\documentclass[reqno]{amsart}
\usepackage{amsmath,amsthm,amssymb}
\usepackage{latexsym}
\usepackage{eucal}
\newtheorem{theorem}{Theorem}[section]
\newtheorem{lemma}{Lemma}[section]

\theoremstyle{definition}

\def\cal{\mathcal}

\let\Im=\undefined
\DeclareMathOperator{\Im}{Im}

\begin{document}
\title{On the absolutely continuous spectrum of Dirac
operator}
\author{Sergey A. Denisov}
\address{Mathematics, 253-37, Caltech, Pasadena, CA, 91125, USA,
{\rm e-mail: denissov@caltech.edu}} \maketitle

\begin{abstract}
We prove that the massless Dirac operator in $\mathbb{R}^3$ with
long-range potential has an a.c. spectrum which fills the whole
real line. The Dirac operators with matrix-valued potentials are
considered as well.
\end{abstract}

\section{ Introduction}

In this paper, we consider Dirac operator for the massless
particle in the external field generated by the long-range
potential
\begin{equation}
H=-i\alpha \cdot \bigtriangledown +V \label{dirac8}
\end{equation}
 Here
\begin{equation*}
\alpha _{j}=\left(
\begin{array}{cc}
0 & \sigma _{j} \\
\sigma _{j} & 0%
\end{array}%
\right) ,\ \sigma _{1}=\left(
\begin{array}{cc}
0 & 1 \\
1 & 0%
\end{array}%
\right) ,\ \sigma _{2}=\left(
\begin{array}{cc}
0 & -i \\
i & 0%
\end{array}%
\right) ,\ \sigma _{3}=\left(
\begin{array}{cc}
1 & 0 \\
0 & -1%
\end{array}%
\right)
\end{equation*}%
Matrices $\sigma _{j\text{ }}$ are called the Pauli matrices.
Denote the exterior of the unit ball in $\mathbb{R}^3$ by
$\Omega$. We use a notation $\Sigma$ for the unit sphere in
$\mathbb{R}^3$. Consider $H$ in the Hilbert space
$[L^2(\Omega)]^4$. Assume that the elements of a self-adjoint
$4\times 4$ matrix-function $V(x)$ are uniformly bounded in
$\Omega$.
 We also assume that $V$ has certain  canonical form after the spherical change of variables. There are
many meaningful potentials that satisfy these assumptions
\cite{Th}. Consider the self-adjoint operator $\cal{H}$, generated
by the boundary conditions $f_3(x)=f_4(x)=0$ as $x\in \Sigma$. We
introduce a matrix
\begin{equation*}
\beta =\left[
\begin{array}{cccc}
1 & 0 & 0 & 0 \\
0 & 1 & 0 & 0 \\
0 & 0 & -1 & 0 \\
0 & 0 & 0 & -1%
\end{array}%
\right]
\end{equation*}
The main result of the paper is the following statement.

{\bf Theorem \ref{th1}}. {\it  Assume $V(x)=v(x)\beta$, where
$v(x)$ is a real-valued, uniformly bounded, scalar function
satisfying the following condition
\begin{equation}
\int\limits_{x\in \Omega} \frac{v^2(x)}{|x|^2+1} <
 \infty
\label{condition}
\end{equation}
Then, $\sigma_{ac}(\cal{H})=\mathbb{R}$. }

This result can be regarded as a PDE (Partial Differential
Equations) analog of the celebrated results by Szeg\"o on
polynomials orthogonal on the unit circle with $\ell^2$ Verblunsky
parameters \cite{Szego, Simonbook}. If one considers the
power-decaying potentials $v(x): |v(x)|\leq
C(|x|+1)^{-0.5-\varepsilon}$, then condition (\ref{condition}) is
satisfied for any $\varepsilon>0$. In this case, we also obtain an
asymptotics for the Green's function.

For the massless Dirac operator,  theorem \ref{th1} solves Simon's
conjecture \cite{Sim} for Schr\"odinger operators. Under more
conditions on $V$, the spectrum of $\cal{H}$ is purely a.c. on
$\mathbb{R}$ \cite{Yam}. One can easily construct an example when
conditions of the theorem \ref{th1} are satisfied and the rich
singular spectrum occurs. In the one-dimensional case, the first
result on the presence of a.c. spectrum for slowly decaying
potentials is due to M. Krein \cite{Kr}.  See also \cite{dk, MNV,
Den7} for the modern development. The existence of wave operators
for the one-dimensional Dirac operator with square summable
potential was proved in \cite{Den}. An interesting paper
\cite{LNS} discusses Szeg\"o-type inequalities for Schr\"odinger
operators with short-range potentials. We will use some ideas from
\cite{LNS}. For discrete multidimensional Schr\"odinger operator
with random slow decay, the existence of wave operators was proved
by Bourgain \cite{Bour}.

The structure of the paper is as follows. In the second section,
we consider one-dimensional Dirac systems with matrix-valued
potentials. Then, in the third part, we deal with a
multidimensional Dirac operator.

The following notations will be used. $|M|=\sqrt{M^{\ast }M}$
denotes the absolute value of matrix $M$, symbol $1$ will often
stand for the identity matrix or operator. As usual,
$C_{0}^{\infty }(\mathbb{R}^{+})$ denotes the space of infinitely
smooth functions (or vector-functions) with the compact support
inside $(0,\infty )$. $\chi_K(x)$ denotes the characteristic
function of the set $K$. $\langle x,y \rangle$ stands for the
inner product of two vectors $x$ and $y$ in $\mathbb{R}^3$.

\section{One-dimensional Dirac operator with matrix-valued potential}

In this section, we study the one-dimensional case. Let us
consider the Dirac operator in the following form

\begin{equation}
D=\left[
\begin{array}{cc}
0 & -d/dr \\
d/dr & 0%
\end{array}%
\right] +V, \quad V=\left[
\begin{array}{cc}
-b & -a \\
-a & b%
\end{array}%
\right] \label{dirac}
\end{equation}%
where $a(r),b(r)$ are $m\times m$ self-adjoint matrices with
locally integrable entries. This form of a general Dirac
operator is called a canonical form (\cite{LS}, pp.48--50). The elements of the Hilbert space are $%
f=(f_{1},f_{2})^{t}$ with $f_{1(2)}\in \left[
L^{2}(\mathbb{R}^{+})\right] ^{m}$. The boundary condition
$f_{2}(0)=0$ defines the self-adjoint operator $\cal{D}$. We start
with an elementary spectral theory of $\cal{D}$. Consider
solutions of the following equation%
\begin{equation*}
D\left[
\begin{array}{c}
\Phi \\
\Psi%
\end{array}%
\right] =\lambda \left[
\begin{array}{c}
\Phi \\
\Psi%
\end{array}%
\right] ,\Phi (0,\lambda )=1,\Psi (0,\lambda )=0, \lambda\in
\mathbb{C}
\end{equation*}%
For any $f$, consider the generalized Fourier transform
\begin{equation*}
F(\lambda )=\int\limits_{0}^{\infty }\Phi ^{\ast }(r,\lambda
)f_{1}(r)dr+\int\limits_{0}^{\infty }\Psi ^{\ast }(r,\lambda )f_{2}(r)dr
\end{equation*}%
There exists non-decreasing $m\times m$ matrix-function $\sigma
(\lambda ),\, \lambda\in \mathbb{R}$ (spectral matrix-valued
measure) such that (\cite{Sakh1}, p.106)
\begin{equation}
\int\limits_{0}^{\infty }\left(
|f_{1}(r)|^{2}+|f_{2}(r)|^{2}\right) dr =\int\limits_{-\infty
}^{\infty }F^{\ast }(\lambda )d\sigma(\lambda )F(\lambda )
\end{equation}
\begin{equation}
\int\limits_{-\infty }^{\infty }\frac{d\sigma(\lambda )}{1+\lambda
^{2}} <\infty \label{cond2}
\end{equation}
The resolvent kernel $R_{z}(r,s)$ of operator $\cal{D}$ has the
following
form %
\begin{equation}
R_{z}(r,s)=\int\limits_{-\infty }^{\infty }\frac{1}{\lambda -z}\left[
\begin{array}{cc}
\Phi (r,\lambda )d\sigma(\lambda )\Phi ^{\ast }(s,\lambda ) & \Phi
(r,\lambda )d\sigma(\lambda )\Psi ^{\ast }(s,\lambda ) \\ \Psi
(r,\lambda )d\sigma(\lambda )\Phi ^{\ast }(s,\lambda ) & \Psi
(r,\lambda )d\sigma(\lambda )\Psi ^{\ast }(s,\lambda )%
\end{array}%
\right] \label{asp}
\end{equation}%
where the integral is understood in the distributional sense.
Notice that

\begin{equation}
\Im R_{z}(0,0)=\left[
\begin{array}{cc}
\displaystyle \Im \int\limits_{-\infty }^{\infty }
(\lambda-z)^{-1} d\sigma (\lambda) & 0
\\
0 & 0%
\end{array}%
\right]  \label{form2}
\end{equation}
where the integral converges due to (\ref{cond2}). Assume
$a(r)=b(r)=0$, if $r>R$. Then, we can always find the Jost
solution $F(r,\lambda)$:
\begin{equation*}
DF=\lambda F,\,F(r,\lambda )=\left[
\begin{array}{c}
F_{1}(r,\lambda ) \\
F_{2}(r,\lambda )%
\end{array}%
\right] =e^{i\lambda r}\left[
\begin{array}{c}
-i \\
1%
\end{array}%
\right] ,\ \text{for }r>R, \lambda \in \overline{\mathbb{C}^{+}}
\end{equation*}
Notice that $F_2(0,\lambda)$ is an entire matrix-valued function.
It is non-degenerate in $\mathbb{C}^+$. Indeed, otherwise we would
have the non-real eigenvalue for the self-adjoint operator
$\cal{D}$. The matrix
\[
\left[
\begin{array}{cc}
\Phi (r,\lambda ) & F_{1}(r,\lambda ) \\
\Psi (r,\lambda ) & F_{2}(r,\lambda )%
\end{array}%
\right]
\]
is non-degenerate for $r=0,\lambda \in \mathbb{C}^{+}$, therefore,
it is non-degenerate for any $r>0$. Consider the following matrix
\begin{equation*}
Z(r,\lambda )=\left[
\begin{array}{cc}
\Phi(r,\lambda) & F_{1}(r,\lambda) \\ \Psi(r,\lambda) & F_{2}(r,\lambda)%
\end{array}%
\right] ^{-1}= \left[
\begin{array}{cc}
Z_{11}(r,\lambda) & Z_{12}(r,\lambda)\\ Z_{21} (r,\lambda)&Z_{22}
(r,\lambda)
\end{array}
\right]
\end{equation*}
Then, the resolvent of $\cal{D}$ can be written in the following
form
\begin{equation*}
R_{\lambda} \left[
\begin{array}{c}
h_{1} \\
h_{2}%
\end{array}%
\right] =\left[
\begin{array}{c}
\Phi (r,\lambda ) \\
\Psi (r,\lambda )%
\end{array}%
\right] \int\limits_{r}^{\infty }\left( -Z_{11}(s,\lambda
)h_{2}(s)+Z_{12}(s,\lambda )h_{1}(s)\right) ds+
\end{equation*}

\begin{equation*}
\left[
\begin{array}{c}
F_{1}(r,\lambda ) \\
F_{2}(r,\lambda )%
\end{array}%
\right] \int\limits_{0}^{r}\left( Z_{21}(s,\lambda
)h_{2}(s)-Z_{22}(s,\lambda )h_{1}(s)\right) ds, \lambda \in
\mathbb{C}^{+}
\end{equation*}
Consequently, the resolvent kernel can be written in the form%
\begin{equation*}
R_{\lambda}(r,s)=\left\{
\begin{array}{c}
\left[
\begin{array}{cc}
\Phi (r,\lambda )Z_{12}(s,\lambda ) & -\Phi (r,\lambda )Z_{11}(s,\lambda )
\\
\Psi (r,\lambda )Z_{12}(s,\lambda ) & -\Psi (r,\lambda )Z_{11}(s,\lambda )%
\end{array}%
\right] ,\text{ if }r<s \\
\\
\left[
\begin{array}{cc}
-F_{1}(r,\lambda )Z_{22}(s,\lambda ) & F_{1}(r,\lambda )Z_{21}(s,\lambda )
\\
-F_{2}(r,\lambda )Z_{22}(s,\lambda ) & F_{2}(r,\lambda )Z_{21}(s,\lambda )%
\end{array}%
\right] ,\text{ if }r>s%
\end{array}%
\right.
\end{equation*}%
In the free case ($a=b=0$), $\Phi(r,\lambda)=\cos(\lambda r)$,
$\Psi(r,\lambda)=-\sin(\lambda r)$,
$F_1(r,\lambda)=-i\exp(i\lambda r)$, $F_2(r,\lambda)=\exp(i\lambda
r)$, $\sigma(\lambda)=d\lambda/\pi$, and
\begin{equation*}
R_{\lambda}^0(r,s)=\left\{
\begin{array}{c}
\left[
\begin{array}{cc}
i\cos(\lambda r) \exp(i\lambda s) & -\cos(\lambda r) \exp(i\lambda
s)
\\
-i\sin (\lambda r) \exp(i\lambda s) & \sin (\lambda r)\exp(i\lambda s)%
\end{array}%
\right] ,\text{ if }r<s \\
\\
\left[
\begin{array}{cc}
i\exp(i\lambda r) \cos(\lambda s) & -i\exp(i\lambda r)
\sin(\lambda s)
\\
-\exp(i\lambda r) \cos(\lambda s) & \exp(i\lambda r)\sin(\lambda s)%
\end{array}%
\right] ,\text{ if }r>s%
\end{array}%
\right.
\end{equation*}
Notice that $Z(0,\lambda )=\left[
\begin{array}{cc}
1 & -F_{1}(0,\lambda )F_{2}^{-1}(0,\lambda ) \\
0 & F_{2}^{-1}(0,\lambda )%
\end{array}%
\right] $ and
\[
\Im R_{\lambda }(0,0) =\frac{1}{2i}\left[ R_{\lambda
}(0,0)-R_{\lambda }^{\ast }(0,0)\right]=
\]
\begin{equation}
=\frac{1}{2i}\left[
\begin{array}{cc}
F_{2}^{\ast -1}(0,\lambda )F_{1}^{\ast }(0,\lambda )-F_{1}(0,\lambda
)F_{2}^{-1}(0,\lambda ) & 0 \\
0 & 0%
\end{array}%
\right],\, \lambda\in \mathbb{C}^+ \label{numer7}
\end{equation}
For $\lambda \in \mathbb{C}$, let us introduce the following
functions: $f_1(r,\lambda)=\exp(-i\lambda r) F_1(r,\lambda)$,\\
\mbox{$f_2(r,\lambda)=\exp(-i\lambda r) F_2(r,\lambda)$}.
Functions $f_1$ and $f_2$ satisfy the following equations
\[
\left\{
\begin{array}{cc}
f_1'=-i\lambda f_1+af_1+\lambda f_2-bf_2\\ f_2'=-\lambda
f_1-bf_1-i\lambda f_2-af_2
\end{array}
\right.
\]
Therefore, a simple algebra yields
\begin{equation}
\frac{d}{dr} \left( \frac{f_1^\ast f_2-f_2^\ast f_1}{i}\right)=
-2\Im \lambda\, |f_1+if_2|^2
\end{equation}
Using the asymptotics at infinity, we obtain
\begin{equation}
\frac{f_1^\ast(r,\lambda) f_2(r,\lambda)-f_2^\ast(r,\lambda)
f_1(r,\lambda)}{i} =2+2\Im \lambda \int\limits_r^\infty
|f_1(\rho,\lambda)+if_2(\rho,\lambda)|^2d\rho \label{grog}
\end{equation}

\begin{lemma} The following  relations hold
\begin{equation}
\frac{F_{1}^{\ast }(r,\lambda )F_{2}(r,\lambda )-F_{2}^{\ast
}(r,\lambda )F_{1}(r,\lambda )}{i}=2,\lambda\in \mathbb{R}
\label{energy1}
\end{equation}%
\begin{equation}
\frac{F_{1}^{\ast }(0,\lambda )F_{2}(0,\lambda )-F_{2}^{\ast
}(0,\lambda )F_{1}(0,\lambda )}{i}\geq 2, \lambda\in \mathbb{C}^+
\label{energy2}
\end{equation}%
\label{l1}
\end{lemma}

Proof. It suffices to use (\ref{grog}) and definition of $f_1$ and
$f_2$. $\blacksquare$

Notice that the lemma is wrong if the Dirac operator is not in the
canonical form. (\ref{energy1}) implies that $F_2(0,\lambda)$ is
non-degenerate on $\mathbb{R}$ also.

\textbf{Corollary.} The following relation holds: $\Im R_{\lambda
}(0,0)_{(1,1)}=$ $F_{2}^{\ast -1}(0,\lambda )F_{2}^{-1}(0,\lambda ),$ if $%
\lambda $ is real (see (\ref{numer7})). Therefore, using
(\ref{form2}), we obtain
\begin{equation}
\sigma' (\lambda )= \pi^{-1}F_{2}^{\ast -1}(0,\lambda
)F_{2}^{-1}(0,\lambda )
\end{equation}
The factorization results of this sort are quite common in the
scattering theory. Direct analogs hold for the so-called Krein
systems (factorization via $\Pi $--functions \cite{Kr}), for
one-dimensional  Schr\"{o}dinger operators with matrix-valued
potentials (factorization via Jost functions \cite{Mar}), for
polynomials with matrix-valued coefficients (factorization via
Szeg\"{o} function \cite{Del,AptNik,Simonbook}). Notice that the
Krein system with coefficient $-A(r)$ yields operator $-\cal{D}$
\cite{Kr}.

 Using the representation of resolvent
by the solutions, one gets%
\begin{equation}
R_{\lambda }(r,0)=\left[
\begin{array}{cc}
-F_{1}(r,\lambda )F_{2}^{-1}(0,\lambda ) & 0 \\
-F_{2}(r,\lambda )F_{2}^{-1}(0,\lambda ) & 0%
\end{array}%
\right] ,\text{ for }\lambda \in \overline{\mathbb{C}^{+}}
\label{form1}
\end{equation}%
So, we can find $F_{2}^{-1}(0,\lambda )$ by following a simple
rule that
turns out to be applicable to PDE%
\begin{equation*}
\lim_{r\rightarrow \infty }R_{\lambda }(r,0)e^{-i\lambda r}=\left[
\begin{array}{cc}
iF_{2}^{-1}(0,\lambda ) & 0 \\
-F_{2}^{-1}(0,\lambda ) & 0%
\end{array}%
\right] ,\ \lambda \in \overline{\mathbb{C}^{+}}
\end{equation*}

\begin{lemma} Consider $a(r),b(r)\in C_{0 }^{\infty
}(\mathbb{R}^{+})$. Then, for the function $F_{2}(0,\lambda )$,
the following inequalities hold true
\begin{equation}
\Vert F_{2} (0,iy)\Vert \leq \exp \left[
Cy^{-1}\int\limits_{0}^{\infty }[\Vert a(r)\Vert ^{2}+\Vert
b(r)\Vert ^{2}]dr\right]  \label{est1}
\end{equation}

 The asymptotics of $F_{2}(0,\lambda )$ as $\lambda \in
\overline{\mathbb{C}^{+}},|\lambda |\rightarrow \infty $ is

\begin{equation}
F_{2}(0,\lambda )=1-\frac{1}{2i\lambda }%
\int\limits_{0}^{\infty } |b(r)+ia(r)|^{2} dr+O(|\lambda |%
^{-2} \label{second}
\end{equation}
\label{lemma2}
\end{lemma}
Proof. Consider a matrix

\begin{equation*}
J= \left[
\begin{array}{cc}
0 & -1 \\
1 & 0%
\end{array}
\right]
\end{equation*}
Upon multiplying by $J$, the equation $DY=\lambda Y$ can be
rewritten as

\begin{equation}
Y'+QY=-\lambda JY,Q=-JV  \label{int2}
\end{equation}
Take

\begin{equation*}
Y_{0 }=\left[
\begin{array}{cc}
-ie^{i\lambda r} & e^{-i\lambda r} \\
e^{i\lambda r} & -ie^{-i\lambda r}%
\end{array}%
\right]
\end{equation*}%
and find the solution to (\ref{int2}) in the following form
$Y=Y_{0 }S$.

Thus
\begin{equation}
S'=-Y_{0 }^{-1}QY_{0 }S=\left[
\begin{array}{cc}
0 & (-b+ia)e^{-2i\lambda r} \\
(-b-ia)e^{2i\lambda r} & 0%
\end{array}%
\right] S  \label{eq1}
\end{equation}
Imposing condition
\begin{equation*}
S(\infty ,\lambda )=\left[
\begin{array}{c}
1 \\
0%
\end{array}%
\right]
\end{equation*}%
on the $2m \times m$ matrix

\begin{equation*}
S=\left[
\begin{array}{c}
S_{1} \\
S_{2}%
\end{array}%
\right]
\end{equation*}
we get $F=Y_{0 }S$. System (\ref{eq1}) can be rewritten in the
following form

\begin{equation}
S_{1}(r,\lambda )=1+\int\limits_{r}^{\infty }e^{-2i\lambda
s}(b(s)-ia(s))S_{2}(s,\lambda )ds
\end{equation}

\begin{equation}
S_{2}(r,\lambda )=\int\limits_{r}^{\infty }e^{2i\lambda
s}(b(s)+ia(s))S_{1}(s,\lambda )ds
\end{equation}
These two equations are easy to iterate. We obtain the following
integral equation for $S_1$

\begin{equation}
S_{1}(r,\lambda )=1+\int\limits_{r}^{\infty }e^{-2i\lambda
s}(b(s)-ia(s))\int\limits_{s}^{\infty }e^{2i\lambda
t}(b(t)+ia(t))S_{1}(t,\lambda )dtds  \label{integ}
\end{equation}
One needs to use Gronwall's inequality to obtain the first
statement of the
lemma. Iterating (\ref{integ}) and integrating by parts, one proves (\ref%
{second}). $\blacksquare $

\textbf{Remark.} Notice that the usual scattering coefficients $A(\lambda )$
and $B(\lambda )$ can be calculated by the formulas

\begin{equation}
\begin{array}{cc}
\displaystyle A(\lambda )=\frac{1}{2}(F_{2}(0,\lambda
)+iF_{1}(0,\lambda ))=S_{1}(0,\lambda ),\\
\displaystyle B(\lambda
)=\frac{1}{2}(F_{2}(0,\lambda )-iF_{1}(0,\lambda
))=-iS_{2}(0,\lambda ),\lambda \in {\mathbb{C}}
\end{array}%
\label{lab1}
\end{equation}
If $a,b\in C_0^\infty (\mathbb{R}^+)$, then the asymptotics of
$A(\lambda)$ is
\begin{equation}
A(\lambda)=1-\frac{1}{2i\lambda }%
\int\limits_{0}^{\infty }|b(r)+ia(r)|^{2}dr+O(|\lambda | ^{-2}),
\lambda\to\infty, \lambda\in \overline{\mathbb{C}^+} \label{asyma}
\end{equation}
 Using
(\ref{energy1}), we get the conservation law
\begin{equation}
|A(\lambda )|^{2}=1+|B(\lambda )|^{2},\ \lambda \in \mathbb{R}
\label{conserv}
\end{equation}
From (\ref{energy2}),
\begin{equation}
|A(\lambda )|^{2}\geq 1+|B(\lambda )|^{2},\ \lambda \in
\mathbb{C}^+ \label{conserv1}
\end{equation}

Notice that if $\xi \in \mathbb{C}^{m},\ \xi \neq 0,$ then the
function $\ln ||F_{2}^{-1}(0,\lambda )\xi ||$ is subharmonic in
$\mathbb{C}^{+}$. Thus, the following inequality holds
\begin{equation}
\pi \ln ||F_{2}^{-1}(0,iy)\xi ||\leq y\int\limits_{-\infty }^{\infty }\frac{%
\ln ||F_{2}^{-1}(0,\lambda)\xi
||}{\lambda^{2}+y^{2}}d\lambda=\frac{y}{2}\int\limits_{-\infty
}^{\infty }\frac{\ln \left( \pi \sigma'(\lambda)\xi ,\xi \right) }{%
\lambda^{2}+y^{2}}d\lambda  \label{est2}
\end{equation}%
Here we used asymptotics of $F_{2}^{-1}(0,\lambda )$ as $|\lambda
|\rightarrow \infty $ and its continuity in
$\overline{\mathbb{C}^{+}}.$ The following theorem is
straightforward.

\begin{theorem} Assume that $||a(r)||,||b(r)||\in
L^{2}(\mathbb{R}^{+})$, $\xi \in \mathbb{C}^{m}$, $\Vert \xi \Vert
=1$. Then, the following inequality holds
\begin{equation}
y^{2}\int\limits_{-\infty }^{\infty }\frac{\ln (\pi
\sigma'(\lambda)\xi ,\xi
)}{y^{2}+\lambda^2}d\lambda>-C\int\limits_{0}^{\infty }\left[
\Vert a(r)\Vert ^{2}+\Vert b(r)\Vert ^{2}\right] dr \label{rrr}
\end{equation}
\end{theorem}

Proof. Assume $a(r),b(r)\in \mathbb{C}_{0 }^{\infty
}(\mathbb{R}^{+})$ first. Then, the estimate (\ref{est1}) from
lemma \ref{lemma2} implies
\begin{equation}
|F_{2}^{-1}(0,iy)|\geq \exp \left( -Cy^{-1}\int\limits_{0}^{\infty
}\left[ \Vert a(r)\Vert ^{2}+\Vert b(r)\Vert ^{2}\right] dr\right)
\label{est3}
\end{equation}%
So, (\ref{est2}) gives the needed estimate. Now, consider any $%
||a(r)||,||b(r)||\in L^{2}(\mathbb{R}^{+})$. We can find the
sequence $a_{n}(r),b_{n}(r)\in \mathbb{C}_0^\infty (\mathbb{R}^+)$
such that
\[
\int\limits_{0}^{\infty }(\Vert a_{n}-a\Vert ^{2}+\Vert
b_{n}-b\Vert ^{2})dr\rightarrow 0
\]
as $n\rightarrow \infty$. Let $\sigma_{n}$ denote the spectral
measure for $a_{n}, b_{n}$.  Then, the second resolvent
identity yields the weak convergence of measures $\sigma_{n}$ to $%
\sigma$. For each individual $\sigma_{n}$, we have the estimate (%
\ref{rrr}). Then, using the standard argument involving semicontinuity of
the entropy \cite{KS}, one gets (\ref{est3}). $\blacksquare $

\textbf{Remark.} Constants in the inequality (\ref{rrr}) do not depend on $m$%
, the size of the matrix. Therefore, one can easily prove an
analogous result for the Dirac operator with square summable
operator-valued potential. As a simple corollary,
$\sigma_{ac}(\cal{D})=\mathbb{R}$.

\textbf{Remark.} The similar results can be proved for the Krein systems \cite%
{Kr, Sakh2} with matrix-valued or operator-valued coefficients.
Notice that the Krein system with coefficient $-A(r)$ in a
standard way generates the Dirac operator $\cal{D}_g=-\cal{D}$
with $\cal{D}$ introduced by (\ref{dirac}).

Assume that $a(r),b(r)$ are integrable with finite support. Consider a fixed vector $%
\xi \in \mathbb{C}^{m}$. Let us find an element of the Hilbert
space with generalized Fourier transform $(\lambda -iy)^{-1}\xi $.
This function is given by the formula

\begin{equation*}
\left[
\begin{array}{c}
\displaystyle \int\limits_{-\infty }^{\infty }(\lambda
-iy)^{-1}\Phi (r,\lambda )d\sigma (\lambda )\xi \\ \displaystyle
\int\limits_{-\infty }^{\infty }(\lambda -iy)^{-1}\Psi (r,\lambda
)d\sigma
(\lambda )\xi%
\end{array}%
\right]
\end{equation*}
and is equal to $R_{iy}(r,0)[\xi ,0]^{t}$ due to (\ref{asp}). At the same time,  (\ref%
{form1}) says that this vector is equal to
\begin{equation*}
-\left[
\begin{array}{c}
F_{1}(r,iy)F_{2}^{-1}(0,iy)\xi \\
F_{2}(r,iy)F_{2}^{-1}(0,iy)\xi%
\end{array}%
\right]
\end{equation*}
Therefore, the following arguments are valid. \ If the
vector-function
\begin{equation}
 h(r)=(h_{1}(r),h_{2}(r))^{t}\in \mathbb{C}^{2m}\label{refe}
\end{equation}
satisfies equation $Dh=iyh$ and decays at infinity, then
$h_{1}(r)=F_{1}(r,iy)\eta ,h_{2}(r)=F_{2}(r,iy)\eta $, where $\eta
\in \mathbb{C}^{m}$.\ Take $\xi =F_{2}(0,iy)\eta =h_{2}(0)$. The
generalized Fourier transform for the function $h(r)$ is
$-(\lambda -iy)^{-1}h_{2}(0)$. The spectral measure of $h$ is
\begin{equation*}
\sigma(\lambda ,h)=\int\limits_{-\infty }^{\lambda }\frac{d(\sigma
(\lambda)h_{2}(0),h_{2}(0))}{\lambda^2+y^{2}}
\end{equation*}
Using the estimate (\ref{est2}), we have
\begin{equation*}
y\int\limits_{-\infty }^{\infty }\frac{\ln \left[ \pi
(\sigma'(\lambda,h)(\lambda^2+y^{2}))\right]
}{\lambda^2+y^{2}}d\lambda\geq 2\pi \ln
||F_{2}^{-1}(0,iy)h_{2}(0)||=2\pi \ln ||\eta ||
\end{equation*}%
or
\begin{equation}
y\int\limits_{-\infty }^{\infty }\frac{\ln \left[
(\sigma'(\lambda,h)\right] }{\lambda^2+y^{2}}d\lambda \geq
C+2\pi\ln ||\eta || \label{paradigma}
\end{equation}%
where the constant $C$ depends on $y$.
 This inequality will play the crucial role later. Roughly
speaking, it means the following.

{\it  If we found at least one solution having the ``free"
asymptotics at some point in $\mathbb{C}^{+}$, then the entropy of
the spectral measure can be controlled by the ``amplitude", i.e.
$||\eta ||$. }

The definition of the spectral measure gives an equality
\begin{equation*}
\int\limits_{-\infty }^{\infty }\sigma'(\lambda,h)d\lambda=
||h||^2
\end{equation*}%
{\it Therefore, the lower bound for $||\eta ||$ yields the lower
bound for
\begin{equation}
\int\limits_{-\infty }^{\infty }\frac{\ln^{-}\sigma'(\lambda,h)}{%
\lambda^2+y^{2}}d\lambda \label{yaka}
\end{equation}
} Consider $a(r),b(r)\in \mathbb{C}_{0 }^{\infty
}(\mathbb{R}^{+}), ||\xi||=1$. The functions $\ln
|(F^{-1}(0,\lambda)\xi,\xi)|$, $\ln |(A^{-1}(\lambda)\xi,\xi)|$,
$\ln\|A(\lambda)\|$ are subharmonic in $\mathbb{C}^+$ and tend to
$0$ as $|\lambda|\to\infty, \lambda\in \overline{\mathbb{C}^+}$.
Write the corresponding inequality at $\lambda=iy$. Taking
$y\to\infty$, one gets the following estimates by comparing the
coefficients against $y^{-1}$.

\begin{eqnarray*}
\int\limits_{-\infty }^{\infty }\ln (\pi \sigma'(\lambda)\xi ,\xi
)d\lambda =2\int\limits_{-\infty }^{\infty }\ln
\|F_2^{-1}(0,\lambda) \xi  \|d\lambda \geq \\ \geq
2\int\limits_{-\infty }^{\infty }\ln |(F_2^{-1}(0,\lambda)
\xi,\xi)| d\lambda
>-C\int\limits_{0}^{\infty } \left[ ||b(r)\xi +i a(r) \xi ||^2
\right] dr; && \\ \int\limits_{-\infty }^{\infty }\ln
|(A^{-1}(\lambda)\xi,\xi)|d\lambda
>-C\int\limits_{0}^{\infty } \left[ ||b(r)\xi +i a(r)\xi||^2 \right] dr; && \\
\int\limits_{-\infty}^{\infty} \ln ||A^{-1}(\lambda)||
d\lambda>-C\int\limits_0^\infty [\|a(r)\|^2+\|b(r)\|^2] dr
\end{eqnarray*}
Notice that estimates $||A^{-1}||\leq 1,$ $||A^{\ast -1}||\leq 1$
imply
\begin{equation}
\begin{array}{ccc}
\displaystyle \int\limits_\Delta \ln
|(A^{-1}(\lambda)\xi,\xi)|d\lambda
>-C\int\limits_{0}^{\infty } \left[ ||b(r)\xi +i a(r)\xi||^2 \right] dr &&
\\
\displaystyle  \int\limits_\Delta \ln ||A^{-1}(\lambda)||
d\lambda>-C\int\limits_0^\infty [\|a(r)\|^2+\|b(r)\|^2] dr
\end{array}
\label{ter}
\end{equation}
for any $\Delta \subset \mathbb{R}$. Consider Dirac operator
(\ref{dirac}). Fix some $\xi \in \mathbb{C}^{m},\left\Vert \xi
\right\Vert =1$. Estimate (\ref{ter}) suggests that condition
$(b(r)+ia(r))\xi \in L^{2}(\mathbb{R}^{+})$ guarantees $\sigma
_{ac}(\cal{D})=\mathbb{R}$.

\begin{theorem}
Let $a(r)$ and $b(r)$
 have locally summable entries. Assume that there is a fixed vector $\xi\in \mathbb{C}^m$ such that $(b+ia)\xi\in
 L^2(\mathbb{R}^+)$.
Then, $\sigma _{ac}(\cal{D})=\mathbb{R}.$ \label{th5}
\end{theorem}

Proof. Without loss of generality, assume $\xi=e_1=(1,0,\ldots,
0)^t$.  Let us start with the compactly supported $a$ and $b$. In
this case, the Jost solution $F(r,\lambda)$ makes sense. Consider
two functions $f_1(\lambda)=A(\lambda)F_2^{-1}(0,\lambda)e_1$ and
$f_2(\lambda)=B(\lambda)F_2^{-1}(0,\lambda)e_1$. From
(\ref{conserv}),
$\|f_1(\lambda)\|^2=\|F_2^{-1}(0,\lambda)e_1\|^2+\|f_2(\lambda)\|^2$
for real $\lambda$. Since $f_1+f_2=e_1$ (by (\ref{lab1})), we have
$|(f_1(\lambda),e_1)|^2=\|F_2^{-1}(0,\lambda)e_1\|^2+|(f_2(\lambda),e_1)|^2
\geq |(F_2^{-1}(0,\lambda)e_1,e_1)|^2+|(f_2(\lambda),e_1)|^2$.
Consider
\[
\alpha(\lambda)=\frac{(f_1(\lambda),e_1)}{(F_2^{-1}(0,\lambda)e_1,e_1)},
\beta(\lambda)=\frac{(f_2(\lambda),e_1)}{(F_2^{-1}(0,\lambda)e_1,e_1)}
\]
Then,
\[
\alpha(\lambda)+\beta(\lambda)=\frac{1}{(F_2^{-1}(0,\lambda)e_1,e_1)}
\]
and
\[
|\alpha(\lambda)|^2\geq 1+|\beta(\lambda)|^2, \lambda \in
\mathbb{R}
\]
Therefore,
\[
|(F_2^{-1}(0,\lambda)e_1,e_1)|\geq \frac{1}{2|\alpha(\lambda)|},
\lambda \in \mathbb{R}
\]
Function $(F_2^{-1}(0,\lambda)e_1,e_1)$ is analytic in
$\overline{\mathbb{C}^+}$ and tends to $1$ as
$|\lambda|\to\infty$, $\lambda\in \overline{\mathbb{C}^+}$. The
numerator of $\alpha(\lambda)$, function $(f_1(\lambda),e_1)$,
does not have zeroes in $\overline{\mathbb{C}^+}$. Indeed, in
$\overline{\mathbb{C}^+}$, we have $\|f_1(\lambda)\|^2\geq
\|F_2^{-1}(0,\lambda)e_1\|^2+\|f_2(\lambda)\|^2$ (by
(\ref{conserv1})) and $f_1(\lambda)+f_2(\lambda)=e_1$. Thus,
$|(f_1(\lambda),e_1)|^2\geq \|F_2^{-1}(0,\lambda)e_1\|^2+
|(f_2(\lambda),e_1)|^2\geq \|F_2^{-1}(0,\lambda)e_1\|^2>0$. So,
the function $\ln|\alpha(\lambda)|$ is superharmonic and
nonnegative on $\mathbb{R}$. For any $\Delta\subset \mathbb{R}$,
we have the estimates
\begin{equation}
\int\limits_\Delta \ln [\pi(\sigma'(\lambda)e_1,e_1)] d\lambda
\geq 2\int\limits_\Delta \ln \|F^{-1}(0,\lambda)e_1\|d\lambda\geq
2\int\limits_\Delta \ln |(F^{-1}(0,\lambda)e_1,e_1)|d\lambda
\label{estt1}
\end{equation}
\[
\geq -2\int\limits_\Delta \ln|\alpha(\lambda)|d\lambda-2\,
|\Delta|\, \ln2
\]
Superharmonicity of $\ln|\alpha(\lambda)|$ implies
\begin{equation}
\pi \ln|\alpha(iy)|\geq y\int\limits_{-\infty}^\infty
\frac{\ln|\alpha(\lambda)|}{\lambda^2+y^2}d\lambda \label{esti5}
\end{equation}
Consider $a_\varepsilon, b_\varepsilon \in C_0^\infty
(\mathbb{R}^+)$ that approximate $a, b$ in $L^2(\mathbb{R}^+)$
norm as $\varepsilon\to 0$. Denote the corresponding spectral
measure by $\sigma_\varepsilon$. Apply (\ref{estt1}) and
(\ref{esti5}). Take $y\to+\infty$ in (\ref{esti5}) and use lemma
\ref{lemma2} and (\ref{asyma}). We obtain
\begin{equation}
\int\limits_{-\infty}^\infty \ln|\alpha_\varepsilon(\lambda)|
d\lambda\leq \frac{\pi}{2} \int\limits_0^\infty
\|(b_\varepsilon(r)+ia_\varepsilon(r))e_1\|^2dr \label{estt2}
\end{equation}
Estimates (\ref{estt1}) and (\ref{estt2}) yield
\begin{equation}
\int\limits_\Delta \ln (\sigma_\varepsilon'(\lambda)e_1,e_1)
d\lambda\geq C_1+C_2\int\limits_0^\infty
\|(b_\varepsilon(r)+ia_\varepsilon(r))e_1\|^2dr
\end{equation}
Since $d\sigma_\varepsilon$ converges weakly to $d\sigma$, we have
\begin{equation}
\int\limits_\Delta \ln (\sigma'(\lambda)e_1,e_1) d\lambda\geq
C_1+C_2\int\limits_0^\infty \|(b(r)+ia(r))e_1\|^2dr \label{estt3}
\end{equation}
Now, take arbitrary $a(r)$ and $b(r)$. Consider truncations
$a_n(r)=a(r)\chi_{[0,n]}(r)$, $b_n(r)=b(r)\chi_{[0,n]}(r)$. Since
the functions $a_n$ and $b_n$ are compactly supported, estimate
(\ref{estt3}) holds for the corresponding measure
$\sigma_n(\lambda)$. It is the general fact of the spectral
theory, that $\sigma_n$ converges to $\sigma$ weakly as
$n\to\infty$. Take $n\to\infty$ to obtain
\begin{equation}
\int\limits_\Delta \ln (\sigma'(\lambda)e_1,e_1) d\lambda\geq
C_1+C_2\int\limits_0^\infty \|(b(r)+ia(r))e_1\|^2dr \label{estt4}
\end{equation}
That completes the proof. $\blacksquare $

Since the constants $C_1$ and $C_2$ in $(\ref{estt4})$ are
independent of $m$, the size of the matrices, the theorem is true
for the  operator-valued Dirac systems as well. Its statement is
very strong, it proves a certain rigidity of the Dirac operator
(\ref{dirac})\footnote{This rigidity is also present for
polynomials orthogonal on the unit circle.}. That makes theorem
\ref{th5} applicable even to some PDE.

\section{Multidimensional Dirac operator}

 In this section, we consider two operators%
\begin{equation*}
H=-i\alpha \cdot \bigtriangledown +V,\ H_{s}=-i\alpha \cdot \bigtriangledown
+V
\end{equation*}%
The first one -- on $\left[ L^{2}(\Omega)\right] ^{4} $ with
boundary conditions $f_{3}=f_{4}=0$ on $\Sigma$. The second
operator -- on $\left[ L^{2}(\mathbb{R}^{3})\right] ^{4}$.
Potential $V(x)$ is always assumed to be symmetric $4\times 4$
matrix with the uniformly bounded entries. For $H$, it is given on
$\Omega$, for $H_s$ -- on $\mathbb{R}^3$. Thus, we have two
self-adjoint operators $\cal{H}$ and $\cal{H}_s$. We also assume
that after the ``spherical change of variables"\footnote{We will
define this change of variables later.}, the
matrix of $V$ has the canonical form in a sense of section 2. The typical example of such a potential is $%
V(x)=v(x)\beta $, where $v(x)$ is a scalar real-valued function (see \cite%
{Th}, p.108 for other interactions and physical explanations). For
simplicity, we will deal with this type of potentials only.

 For $\alpha ,\beta $, we have
\begin{equation*}
\alpha _{k}\alpha _{l}+\alpha _{l}\alpha _{k}=2\delta _{kl},\
\alpha _{k}\beta +\beta \alpha _{k}=0,\ \beta ^{2}=1;\ k,l=1,2,3
\end{equation*}%
Notice that by letting $I=-i\sigma _{1},J=-i\sigma _{2},K=-i\sigma _{3},$ we
have relations
\begin{equation*}
I^{2}=J^{2}=\lambda^2=-1,
\end{equation*}%
\begin{equation*}
IJ=-JI,\ IK=-KI,\ KJ=-JK,\ IJ=K,\ JK=I,\ KI=J
\end{equation*}%
which makes $I,J,$ and $K$ quaternions. The following algebraic
relations are easy to verify.

\begin{lemma} If $\gamma \in \Sigma$, then
\[
\begin{array}{l}
(\alpha _{1}\gamma _{1}+\alpha _{2}\gamma _{2}+\alpha _{3}\gamma
_{3})^{2} =1 \\ (\alpha _{1}\gamma _{1}+\alpha _{2}\gamma
_{2}+\alpha _{3}\gamma _{3}+1)^{2} =2(\alpha _{1}\gamma
_{1}+\alpha _{2}\gamma _{2}+\alpha _{3}\gamma _{3}+1)
\\
(\alpha _{1}\gamma _{1}+\alpha _{2}\gamma _{2}+\alpha _{3}\gamma
_{3}+1)\beta (\alpha _{1}\gamma _{1}+\alpha _{2}\gamma _{2}+\alpha
_{3}\gamma _{3}+1) =0
\end{array}
\]
\label{lll1}
\end{lemma}

\begin{lemma} The resolvent kernel of the free Dirac
operator $\cal{H}_s^0 $ has the following form (\cite{Th}, p. 39)
\begin{equation}
G_{\lambda }^{0 }(x,s)=\left( i\frac{\alpha \cdot (x-s)}{|x-s|^{2}}%
+\lambda \frac{\alpha \cdot (x-s)}{|x-s|}+\lambda \right)
\frac{e^{i\lambda |x-s|}}{4\pi |x-s|} \label{fri}
\end{equation}%
\end{lemma}
Proof. Use the identities
\begin{equation}
(\cal{H}_s^0-\lambda )^{-1}=(\cal{H}_s^0+\lambda
)({\cal{H}_s^0}^{2}-\lambda ^{2})^{-1},\
{\cal{H}_{s}^0}^{2}=-\Delta \label{ident1}
\end{equation}
 and
the explicit formula for the resolvent kernel of the free
Laplacian.

We will start with the following theorem.

\begin{theorem} Assume $V(x)=v(x)\beta$, where $v(x)$ is a
real-valued, uniformly bounded, scalar function satisfying the
following condition
\begin{equation}
\int\limits_{x\in \Omega} \frac{v^2(x)}{|x|^2+1} \leq \infty
\label{condition1}
\end{equation}
Then, $\sigma_{ac}(\cal{H})=\mathbb{R}$.\label{th1}
\end{theorem}

Proof. As in \cite{LNS}, we can assume without loss of generality,
that $v(x)=0$ in \mbox{$1<|x|<2$}. Indeed, otherwise we subtract
function $v(x)\beta \chi_{1<|x|<2}(x)$ from $v(x)\beta$. The
resolvent of $\cal{H}^0$ is an integral operator. One can obtain
an expression for its kernel by using identity (\ref{ident1}) with
$\cal{H}^0$ instead of $\cal{H}_s^0$. Therefore, the resolvent of
$\cal{H}$ is an integral operator too and the trace-class argument
\cite{RS} can be applied. Now, we take any nontrivial infinitely
smooth radially-symmetric function $f(x)$ with support in
$\{1<|x|<2\}$. We will show that the spectral measure of the
element
\begin{equation}
f(x)=(f(r),0,0,0)^t \label{chika}
\end{equation} has an a.c. component which
support fills $\mathbb{R}$. The proof consists of two steps. In
the first step, we represent operator $\cal{H}$ in the canonical
form (\ref{dirac}) with unbounded operator-valued coefficients.
Then, we apply the results of section 1 to obtain the needed
estimates for the entropy of the spectral measure of ${f}$.

 Let us start with the suitable change of variables.
Consider the Dirac operator $\cal{H}$. Let us write this operator
in the spherical coordinates (see \cite{Th}, p.126 or
\cite{Weidmann}, p.16). The standard unitary operator
$f(x)\in L^{2}(\Omega)\overset{\cal{U}}{\rightarrow }%
F(r)=rf(r\sigma ),\sigma \in \Sigma$ maps scalar functions of
three
variables to a vector-valued function of one variable.\ For almost any $r>1$, $%
F(r)\in L^{2}(\Sigma)$. Following \cite{Weidmann}, we consider the
following  auxiliary operators
\[
\sigma_0= \left[ \begin{array}{cc} 1&0\\0&1
\end{array}\right],
\,\sigma_r=r^{-1}\sum_{j=1}^3 \sigma_j x_j, \,
p_j=-i\partial/\partial x_j,\,
\]
\[
 p_r=-i(\partial/\partial
r+r^{-1}),\, L_j=x_{j+1}p_{j+2}-x_{j+2}p_{j+1},\,
s=\sigma_0+\sum_{j=1}^3 \sigma_j L_j \] where indices $j$ are
understood modulo $3$, as usual. Notice that the operator $p_r$ is
unitary to $-id/dr$ under $\cal{U}$ and $s$ is independent of $r$.
We want to get a matrix representation of $\cal{H}$ convenient for
us. To do that, we take a unitary, independent or $r$ matrix $U$
\[
U= \left[
\begin{array}{cc}
\sigma_0 & 0\\ 0 & -i\sigma_r
\end{array}
\right]
\]
Then, (\cite{Weidmann}, p. 18)
\begin{equation}
U^{\ast}\cal{H}U= \left[
\begin{array}{cc}
0 & -i\sigma_0\\ i\sigma_0 & 0
\end{array}
\right]p_r -\frac{1}{r} \left[
\begin{array}{cc}
0 & s\\ s & 0
\end{array}
\right] +v(x)U^{\ast}\beta U
\end{equation}
Operator $s$ is selfadjoint in $[L^2(\Sigma)]^2$. Denote its
orthonormal eigenfunctions by $\psi_{(n)}$ where $n$ is a
multiindex. Therefore, in
 $\left[ L^{2}(\Sigma)\right] ^{4}$, we can introduce an orthonormal basis spanned by the functions
\[
\Psi^+_{(n)}=
\left[
\begin{array}{c}
\psi_{(n)}\\ 0
\end{array}
\right],
\Psi^-_{(k)}=
\left[
\begin{array}{c}
0\\ \psi_{(k)}
\end{array}
\right]
\]
Then, any function $f(x)\in \left[ L^{2}(\Omega) \right] ^{4}$ can
be represented as a sum $f(x)\overset{\cal{U}}{\sim}
F(r)=\sum\limits_{n}\varphi _{(n)}^{+}(r)\Psi
_{(n)}^{+}+\sum\limits_{k}\varphi _{(k)}^{-}(r)\Psi _{(k)}^{-}$.
The matrix of the unperturbed operator $\cal{H}_{0 }$ can be
written as (\cite{Th}, p.128 and \cite{Weidmann}, p.22)
\begin{equation*}
\left[
\begin{array}{cccccc}
0 & 0 & \ldots & -\frac{d}{dr}-\frac{\kappa _{(1)}}{r} & 0 &
\ldots
\\ 0 & 0 & \ldots & 0 & -\frac{d}{dr}-\frac{\kappa _{(2)}}{r} &
\ldots \\ \ldots & \ldots & \ldots & \ldots & \ldots & \ldots \\
\frac{d}{dr}-\frac{\kappa _{(1)}}{r} & 0 & \ldots & 0 & 0 & \ldots
\\ 0 & \frac{d}{dr}-\frac{\kappa _{(2)}}{r} & \ldots & 0 & 0 &
\ldots \\
\ldots & \ldots & \ldots & \ldots & \ldots & \ldots%
\end{array}%
\right] \left[
\begin{array}{c}
\varphi _{(1)}^{+}(r) \\ \varphi _{(2)}^{+}(r) \\ \ldots \\
\varphi _{(1)}^{-}(r) \\ \varphi _{(2)}^{-}(r) \\
\ldots%
\end{array}%
\right]
\end{equation*}%
where $\kappa_{(n)}$ are eigenvalues of $s$. The boundary
conditions at $r=1$ are $\varphi _{(n)}^{-}(1)=0$.

Now, we want to work with the new Hilbert space $\cal{L}$, the
$L^2$ space of vector-functions
$\Phi(r)=(\varphi_{(1)}^+(r),\ldots,\varphi_{(1)}^-(r),\ldots)^t$,
with the norm given by the formula below
\[
\|\Phi(r)\|^2=\int\limits_1^\infty \sum_n \left[
|\varphi_{(n)}^+(r)|^2+|\varphi_{(n)}^-(r)|^2 \right] dr<\infty
\]
So, the free multidimensional Dirac operator can be represented as
an infinite orthogonal sum of one-dimensional Dirac operators.
Using elementary properties of the Pauli matrices, we get
\[
v(x)U^{\ast}\beta U=v(x)\beta
\]
Consider the multiplication by $v(x)$ in $[L^2(\Sigma)]^2$ as a
self-adjoint bounded operator with matrix $-b(r)$,
$b(r)=\{b_{ij}(r)\}, i,j=1,\ldots, \infty $. Since the section 2
deals with an interval $[0,\infty)$ rather than $[1,\infty)$, we
shift the argument $r$ by one. We end up with the canonical
representation (\ref{dirac}), where $a(r)$ is an unbounded
operator in the diagonal form for each $r>0$. Notice that the
constant function $(1,0)^t\in [L^2(\Sigma)]^2$ is an eigenfunction
of the operator $s$ corresponding to the eigenvalue $1$. For
$\Psi_{(2)}^+$ and $\Psi_{(1)}^-$, we choose the normed vectors
collinear to $(1,0,0,0)^t$ and $(0,0,1,0)^t$, respectively. We
have $\kappa_{(1)}=1$ and the function $f$, given by
(\ref{chika}), corresponds to
\mbox{$\Phi(r)=((r+1)f(r+1),0,\ldots,0,\ldots)$} in the
vector-valued representation. From now on, we will deal with the
representation (\ref{dirac}) of an operator $\cal{H}$.

In the second step of the proof, we implement the main result of
the second section, theorem \ref{th5}. We can not do that
directly, because we are dealing with an operator-valued
coefficient $a(r)$ unbounded for each $r>0$. To avoid this
difficulty, we consider operators $\cal{D}_n$ with $b_n=P_n b
P_n$, where $P_n$ denotes the orthogonal projection onto the
linear combination of the first $n$ functions $\psi_{(k)}$. The
matrix of $b_n(r)$ is
\begin{equation*}
b_n (r)=\left[
\begin{array}{ccccc}
b_{11}(r) & \cdots & b_{1n}(r) & 0 & \cdots \\ \cdots & \cdots &
\cdots & \cdots & \cdots \\ b_{n1}(r) & \cdots & b_{nn}(r) & 0 &
\cdots \\ 0 & \cdots & 0 & 0 & \cdots \\
\cdots & \cdots & \cdots & \cdots & \cdots%
\end{array}
\right], r>0
\end{equation*}%
Each of $\cal{D}_n$ can be written as an orthogonal sum of the
Dirac operator $\cal{\hat{D}}_n$ with matrix-valued potential
$V_n$ of size $n\times n$ and an infinite number of the
one-dimensional Dirac operators with scalar potentials. Denote by
$\cal{L}_n$ the subspace of $\cal{L}$ on which $\hat{\cal{D}}_n$
acts.
 Matrix
$V_n$ has the following form
\[
V_n= \left[
\begin{array}{cc}
-b_n(r) & -a_n(r) \\ -a_n(r) & b_n(r)
\end{array}
\right], r>0
\]
with
\[
a_n (r)=\left[
\begin{array}{cccc}
\frac{\kappa_{(1)}}{r+1} & 0 & \cdots &  0 \\
 \cdots & \cdots & \cdots & \cdots \\
 0 & \cdots & 0 & \frac{\kappa_{(n)}}{r+1}
\end{array}
\right]
\]
Notice that the function ${\Phi(r)}$ lies in $\cal{L}_n$ for any
$n$. Denote the spectral matrix-valued measure of
$\cal{\hat{D}}_n$ by $\sigma_n(\lambda)$.  Since $b(r)=0$ on
$[0,1]$, the spectral measure $d\mu_n(\lambda)$ of $\Phi(r)$ is
equal to $|\rho(\lambda)|^2 (d\sigma_n(\lambda)e_1,e_1)$, where
\[
\rho(\lambda)=\int\limits_0^1 (r+1)f(r+1)\left[\cos(\lambda
r)+\frac{\sin(\lambda r)}{\lambda}\right] dr
\]
Here,
\[
 \cos(\lambda r)+\frac{\sin(\lambda r)}{\lambda},\quad
\frac{r\cos(\lambda r)}{\lambda (r+1)}-\sin(\lambda
r)\left(1+\frac{1}{\lambda^2(r+1)}\right)
\]
are generalized eigenfunctions for the Dirac operator
(\ref{dirac}) with potential
\[
V=\left[
\begin{array}{cc}
 0 & \displaystyle -\frac{1}{r+1} \\
\displaystyle -\frac{1}{r+1} & 0%
\end{array}%
\right]
\]
From the proof of theorem \ref{th5} (estimate (\ref{estt3})), we
get
\begin{equation}
\int\limits_\Delta \ln (\sigma_n'(\lambda)e_1,e_1) d\lambda\geq
C_1+C_2\int\limits_0^\infty \|(b_n(r)+ia_n(r))e_1\|^2dr
\end{equation}
for any finite interval $\Delta\in \mathbb{R}$.  Notice that
\[
\int\limits_0^\infty \|a_n(r)e_1\|^2 dr=\int\limits_0^\infty
(r+1)^{-2}dr
\]
and
\[
\|b_n(r)e_1\|^2 \leq \|b(r)e_1\|^2=\int\limits_{\tau\in\Sigma}
v^2((r+1)\tau)d\tau
\]
So,
\[
\int\limits_0^\infty \|b_n(r)e_1\|^2dr \leq \int\limits_0^\infty
\int\limits_{\tau\in\Sigma} v^2((r+1)\tau) d\tau dr\leq C
\int\limits_{\Omega} \frac{v^2(x)}{|x|^2+1}dx
\]
and we clearly have
\begin{equation}
\int\limits_\Delta \ln \mu_n'(\lambda) d\lambda\geq C>-\infty
\end{equation}
with some constant $C$ independent of $n$. Notice that $b_n\to b$
strongly. So, $d\mu_n(\lambda)$ converges weakly to $d\mu$, the
spectral measure of $f$ with respect to the initial operator
$\cal{H}$. The semicontinuity of the entropy \cite{KS} implies
\[
\int\limits_\Delta \ln \mu'(\lambda)d\lambda>C>-\infty
\]
  $\blacksquare$

The reduction of $\cal{H}$ to a one-dimensional system with the
operator-valued potential is, probably, not necessary. One could
have introduced the radiative operator and worked with this
operator directly avoiding an approximation by Dirac operators
with matrix-valued potentials (see \cite{Sai}). In the meantime,
this reduction is not too difficult and it shows how very general
facts for the matrix-valued orthogonal systems are applied to
different PDE's.

We do not consider the question of existence of wave operators. It
might be that the problem can be solved by using an approach of
\cite{Den}.

In the next theorem, we establish an asymptotics of the Green's
function for the operator $\cal{H}_s$. For $\cal{H}$, that means
existence of the function $h$ (see (\ref{refe})) satisfying
homogeneous equation $Hh=\lambda h, \lambda \in \mathbb{C}^+$ with
the well-controlled ``amplitude". Due to (\ref{paradigma}) and
(\ref{yaka}), we have a certain estimate on the entropy of the
corresponding spectral measure. That, in particular, gives another
proof of $\sigma_{ac}(\cal{H})=\mathbb{R}$ for the case of a
power-decay.

\begin{theorem} Assume $v(x)$ is given on $\mathbb{R}^3$ and satisfies an estimate
$|v(x)|<C_{v}(|x|+1)^{-0.5-\varepsilon },$ with fixed $\varepsilon
>0$ and $C_{v}$ sufficiently small. Then, the resolvent
kernel $G_{\lambda}(x,0)$ of operator $H_{s}$ at a point
$\lambda=i$ has the following representation
\begin{equation}
G_{i}(x,0)=\frac{e^{-|x|}}{4\pi |x|}\left[ \left( i\frac{\alpha \cdot x}{|x|}%
+i\right) \cal{P}_{1}(x)+(|x|+1)^{-0.5}\cal{P}_{2}(x)\right]
\label{est13}
\end{equation}%
where
\begin{equation}
|\cal{P}_2(x)|<C \label{numer}
\end{equation}
uniformly in $\mathbb{R}^{3}$, $\|\cal{P}_1-1\|<\delta$,
$\delta\to 0$ as $C_v\to 0$.
 Positive constant $C$ depends on $C_{v}$ and $%
\varepsilon $ only. \label{th7}
\end{theorem}

We need the smallness of $C_v$ to guarantee convergence of a
certain series. In general situation, one can take the spectral
parameter $\lambda$ sufficiently far from the spectrum. Or, we can
take $\lambda=i$, and let potential be zero on the large ball
centered at origin. That would make constant $C_v$ as small as we
want\footnote{ \, With respect to a smaller $\varepsilon$.} but
would not change the scattering picture.

Let us prove some auxiliary lemmas first.

\begin{lemma} The following estimate holds $(1<\rho
\leq 2|x|/3)$
\begin{equation}
\int\limits_{|y|=\rho }e^{-|x-y|-|y|}d\tau _{y} <C\rho e^{-|x|}
\label{est111}
\end{equation}
\begin{equation*}
\int\limits_{|y|=\rho }e^{-|x-y|-|y|}\sin \zeta (x,y)d\tau _{y} <C\sqrt{%
\rho }e^{-|x|}
\end{equation*}
\begin{equation*}
\int\limits_{|y|=\rho }e^{-|x-y|-|y|}\sin^2 \zeta (x,y)d\tau _{y}
<C e^{-|x|}
\end{equation*}
 where $\zeta (x,y)$ is an angle
between $x$ and $y$. \label{ll1}
\end{lemma}
Proof. Without loss of generality, assume that $x=(0,0,|x|)$. Introducing
the spherical coordinates $y_{1}=\rho \cos \theta \cos \varphi ,y_{2}=\rho
\cos \theta \sin \varphi ,y_{3}=\rho \sin \theta ,$ we get%
\begin{eqnarray*}
&&\rho ^{2}\int\limits_{-\pi }^{\pi }d\varphi \int\limits_{-\pi
/2}^{\pi /2}d\theta \cos \theta \exp (-\rho -\sqrt{|x|^{2}+\rho
^{2}-2|x|\rho \sin \theta }) \\ &<&C\rho
^{2}e^{-|x|}\int\limits_{-\pi /2}^{\pi /2}d\theta \cos \theta \exp
\left[ -c|x|\rho (|x|-\rho )^{-1}(1-\sin \theta )\right]
<Ce^{-|x|}\frac{|x|-\rho }{|x|}\rho
\end{eqnarray*}%
Estimate (\ref{est111}) is now straightforward. The other statement of the
lemma can be proved similarly.$\blacksquare $

Take any two vectors $x, y\in \mathbb{R}^3$. The following
inequality is obvious
\begin{equation}
\left\vert \frac{y}{|y|}-\frac{x}{|x|}\right\vert <C\sin \zeta
\label{esti1}
\end{equation}
Together with $\zeta(x,y)$, consider $\chi(x,y)$ -- an angle
between $x-y$ and $x$. From the sine-theorem, $\sin \chi
=|y-x|^{-1}|y|\sin \zeta $. Consequently,
\begin{equation}
\left\vert \frac{x-y}{|x-y|}-\frac{x}{|x|}\right\vert <C\sin
\chi=C |y-x|^{-1}|y|\sin \zeta \label{est12}
\end{equation}%
By the triangle inequality,
\begin{equation}
\left| \frac{y}{|y|}-\frac{x-y}{|x-y|}\right|\leq \left|
\frac{y}{|y|}-\frac{x}{|x|}\right|+\left|
\frac{x}{|x|}-\frac{x-y}{|x-y|}\right|
\end{equation}
For $|y|<2|x|/3$,
\begin{equation}
\left| \frac{y}{|y|}-\frac{x-y}{|x-y|}\right|\leq C\sin\zeta
\label{yalo}
\end{equation}
Let $|x|>1$ and $\Upsilon=\{y: |y|>2|x|/3, |x-y|>2|x|/3\}$.
\begin{lemma} The following estimate holds
\begin{equation}
\int\limits_\Upsilon \exp(-|x-y|-|y|)dy \leq C \exp(-\gamma|x|)
\end{equation}
with $\gamma>1$. \label{ll0}
\end{lemma}
Proof. Indeed, in $\Upsilon$,
\[
|x-y|+|y|>|y|/5+16|x|/15
\]
Taking $\gamma=16/15$, we obtain the statement of the lemma.
$\blacksquare$

In the following three lemmas, we will be estimating certain
integrals over the $\mathbb{R}^3$. Lemma \ref{ll0} shows that the
contribution coming from the integration over $\Upsilon$ is small
and can be neglected.

\begin{lemma} The following bound is true
\begin{equation*}
\int_{\mathbb{R}^{3}}\frac{e^{-|x-y|}}{|x-y|^{2}}\frac{e^{-|y|}}{|y|^{1.5+\varepsilon
}+1}dy<C\frac{e^{-|x|}}{|x|^{1.5}+1}
\end{equation*}
\label{dwarf}
\end{lemma}

Proof. By lemma \ref{ll1},
\begin{eqnarray*}
\int_{|y|<2|x|/3}\frac{e^{-|x-y|}}{|x-y|^{2}}\frac{e^{-|y|}}{%
|y|^{1.5+\varepsilon }+1}dy
<C\frac{e^{-|x|}}{|x|^{2}+1}\int\limits_{0}^{|x|}\frac{\rho }{\rho
^{1.5+\varepsilon }+1}d\rho <C\frac{e^{-|x|}}{|x|^{1.5+\varepsilon
}+1}; &&
\\
\int_{|y-x|<2|x|/3}\frac{e^{-|x-y|}}{|x-y|^{2}}\frac{e^{-|y|}}{%
|y|^{1.5+\varepsilon }+1}dy =\int_{|y|<2|x|/3}\frac{e^{-|x-y|}}{|y|^{2}}%
\frac{e^{-|y|}}{|x-y|^{1.5+\varepsilon }+1}dy &&\\
<C\frac{e^{-|x|}}{|x|^{1.5+\varepsilon }+1}\int\limits_{0}^{|x|}\frac{%
d\rho }{\rho +1}<C\frac{e^{-|x|}}{|x|^{1.5}+1}
\end{eqnarray*}%
 $\blacksquare $

\begin{lemma} The following relation is true
\begin{eqnarray*}
&&\int_{\mathbb{R}^{3}}\frac{e^{-|x-y|}}{|x-y|}\left( \frac{\alpha \cdot (x-y)}{|x-y|}%
+1\right) \frac{e^{-|y|}}{|y|^{2+\varepsilon }+1}dy \\
&=&\frac{e^{-|x|}}{|x|+1}\left[ \left( \frac{\alpha \cdot x}{|x|}+1\right)
\varphi _{1}(x)+(|x|+1)^{-0.5}\varphi _{2}(x)\right]
\end{eqnarray*}%
where $\varphi _{1(2)}(x)$ are matrix-functions uniformly bounded
in $\mathbb{R}^{3}$. \label{ork}
\end{lemma}

Proof.
\begin{eqnarray*}
\int_{\mathbb{R}^{3}}\frac{e^{-|x-y|}}{|x-y|}\left( \frac{\alpha \cdot (x-y)}{|x-y|}%
+1\right) \frac{e^{-|y|}}{|y|^{2+\varepsilon }+1}dy && \\
=\int_{\mathbb{R}^{3}}\frac{e^{-|x-y|}}{|x-y|}\left( \frac{\alpha \cdot (x-y)}{|x-y|}-%
\frac{\alpha \cdot x}{|x|}\right) \frac{e^{-|y|}}{|y|^{2+\varepsilon }+1}dy
&& \\
+\left( \frac{\alpha \cdot x}{|x|}+1\right) \int_{\mathbb{R}^{3}}\frac{e^{-|x-y|}}{%
|x-y|}\frac{e^{-|y|}}{|y|^{2+\varepsilon }+1}dy &=&I_{1}+I_{2}
\end{eqnarray*}
Following the proof of lemma \ref{dwarf},  we get
\begin{equation*}
\int_{\mathbb{R}^{3}}\frac{e^{-|x-y|}}{|x-y|}\frac{e^{-|y|}}{|y|^{2+\varepsilon }+1}%
dy<C\frac{e^{-|x|}}{|x|+1}\quad {\rm (sharp!)}
\end{equation*}%
For $I_{1},$ we have
\begin{eqnarray*}
&&\int_{|x-y|<2|x|/3}\frac{e^{-|x-y|}}{|x-y|}\left( \frac{\alpha \cdot (x-y)%
}{|x-y|}-\frac{\alpha \cdot x}{|x|}\right) \frac{e^{-|y|}}{%
|y|^{2+\varepsilon }+1}dy \\
&=&-\int_{|y|<2|x|/3}\frac{e^{-|y|}}{|y|}\left( \frac{\alpha \cdot y}{|y|}-%
\frac{\alpha \cdot x}{|x|}\right) \frac{e^{-|x-y|}}{|x-y|^{2+\varepsilon }+1}%
dy
\end{eqnarray*}%
By (\ref{esti1}) and lemma \ref{ll1},
\begin{eqnarray*}
\int_{|y|<2|x|/3}\frac{e^{-|y|}}{|y|}\left\vert \frac{y}{|y|}-\frac{x}{|x|}%
\right\vert \frac{e^{-|x-y|}}{|x-y|^{2+\varepsilon }+1}dy
<\frac{Ce^{-|x|}}{|x|^{2+\varepsilon }+1}\int\limits_{0}^{|x|}\rho
^{-0.5}d\rho <C\frac{e^{-|x|}}{|x|^{1.5+\varepsilon }+1}
\end{eqnarray*}%
We now estimate integral in $I_{1}$ over $|y|<2|x|/3$. By
(\ref{est12}) and lemma \ref{ll1},
\begin{eqnarray*}
&&\left\vert \int_{|y|<2|x|/3}\frac{e^{-|x-y|}}{|x-y|}\left( \frac{\alpha
\cdot (x-y)}{|x-y|}-\frac{\alpha \cdot x}{|x|}\right) \frac{e^{-|y|}}{%
|y|^{2+\varepsilon }+1}dy\right\vert \\
&<&\frac{Ce^{-|x|}}{|x|^{2}+1}\int\limits_{0}^{|x|}\frac{\rho ^{1.5}}{\rho
^{2+\varepsilon }+1}d\rho <C\frac{e^{-|x|}}{|x|^{1.5+\varepsilon }+1}
\end{eqnarray*}%
$\blacksquare$

\begin{lemma} If $v(x)$ is a real-valued scalar
function satisfying an estimate
\\ $|v(x)|<(|x|+1)^{-0.5-\varepsilon },$ then the following
representation holds
\begin{eqnarray*}
&&\int_{\mathbb{R}^{3}}\frac{e^{-|x-y|}}{|x-y|}\left( \frac{\alpha \cdot (x-y)}{|x-y|}%
+1\right) \beta v(y)\left( \frac{\alpha \cdot y}{|y|}+1\right) \frac{e^{-|y|}%
}{|y|}dy \\
&=&\frac{e^{-|x|}}{|x|+1}\left[ \left( \frac{\alpha \cdot x}{|x|}+1\right)
\varphi _{1}(x)+(|x|+1)^{-0.5}\varphi _{2}(x)\right]
\end{eqnarray*}%
where $\varphi _{1(2)}(x)$ are, again, matrix-functions uniformly
bounded in $\mathbb{R}^{3}$. \label{goblin}
\end{lemma}

Proof. From the lemma \ref{lll1}, we infer
\begin{equation*}
\left( \frac{\alpha \cdot (x-y)}{|x-y|}+1\right) \beta \left( \frac{\alpha
\cdot (x-y)}{|x-y|}+1\right) =0
\end{equation*}%
Therefore,%
\begin{eqnarray*}
\int_{\mathbb{R}^{3}}\frac{e^{-|x-y|}}{|x-y|}\left( \frac{\alpha \cdot (x-y)}{|x-y|}%
+1\right) \beta v(y)\left( \frac{\alpha \cdot y}{|y|}+1\right) \frac{e^{-|y|}%
}{|y|}dy &=& \\
=\int_{\mathbb{R}^{3}}\frac{e^{-|x-y|}}{|x-y|}\left( \frac{\alpha \cdot (x-y)}{|x-y|}%
+1\right) \beta v(y)\left( \frac{\alpha \cdot y}{|y|}-\frac{\alpha
\cdot (x-y)}{|x-y|}\right) \frac{e^{-|y|}}{|y|}dy && \\
=J_{1}+J_{2}&&
\end{eqnarray*}
where
\begin{equation*}
J_{1}=\left( \frac{\alpha \cdot x}{|x|}%
+1\right)\int_{\mathbb{R}^{3}}\frac{e^{-|x-y|}}{|x-y|} \beta
v(y)\left( \frac{\alpha \cdot y}{|y|}-\frac{\alpha \cdot
(x-y)}{|x-y|}\right) \frac{e^{-|y|}}{|y|}dy
\end{equation*}%
and%
\begin{equation*}
J_{2}=\int_{\mathbb{R}^{3}}\frac{e^{-|x-y|}}{|x-y|}\left( \frac{\alpha \cdot (x-y)}{%
|x-y|}-\frac{\alpha \cdot x}{|x|}\right) \beta v(y)\left( \frac{\alpha \cdot
y}{|y|}-\frac{\alpha \cdot (x-y)}{|x-y|}\right) \frac{e^{-|y|}}{|y|}dy
\end{equation*}

Consider $J_1$. Using (\ref{yalo}) and lemma \ref{ll1}, we obtain
\begin{eqnarray*}
&&\int_{|y|<2|x|/3}\frac{e^{-|x-y|}}{|x-y|}|v(y)|\left\vert \frac{y}{|y|}-%
\frac{x-y}{|x-y|}\right\vert \frac{e^{-|y|}}{|y|}dy \\
&<&C\frac{e^{-|x|}}{|x|+1}\int\limits_{0}^{|x|}\frac{\rho
^{0,5}}{\rho ^{1.5+\varepsilon }+1}d\rho <C\frac{e^{-|x|}}{|x|+1}\
\ (\text{sharp!})
\end{eqnarray*}%
\begin{eqnarray*}
\int_{|y-x|<2|x|/3}\frac{e^{-|x-y|}}{|x-y|}|v(y)|\left\vert \frac{y}{|y|}-%
\frac{x-y}{|x-y|}\right\vert \frac{e^{-|y|}}{|y|}dy && \\
=\int_{|y|<2|x|/3}\frac{e^{-|x-y|}}{|x-y|}|v(x-y)|\left\vert \frac{y}{|y|}-%
\frac{x-y}{|x-y|}\right\vert \frac{e^{-|y|}}{|y|}dy && \\
<C\frac{e^{-|x|}}{|x|^{1.5+\varepsilon
}+1}\int\limits_{0}^{|x|}\frac{\rho ^{0.5}}{\rho +1}d\rho
<C\frac{e^{-|x|}}{|x|^{1+\varepsilon }+1} &&
\end{eqnarray*}

Using (\ref{est12}), (\ref{yalo}), and lemma \ref{ll1}, we get the following inequalities to estimate $%
|J_{2}|$
\begin{eqnarray*}
&|J_2|\leq &\int_{|y|<2|x|/3}\frac{e^{-|x-y|}}{|x-y|}\left\vert \frac{x-y}{|x-y|}-%
\frac{x}{|x|}\right\vert |v(y)|\left\vert \frac{y}{|y|}-\frac{x-y}{|x-y|}%
\right\vert \frac{e^{-|y|}}{|y|}dy \\
&<&\frac{Ce^{-|x|}}{|x|^{2}+1}\int\limits_{0}^{|x|}\frac{\rho }{\rho
^{1.5+\varepsilon }+1}d\rho <\frac{Ce^{-|x|}}{|x|^{1.5+\varepsilon }+1}
\end{eqnarray*}%
For the region $|x-y|<2|x|/3$, we have%
\begin{eqnarray*}
\int_{|y-x|<2|x|/3}\frac{e^{-|x-y|}}{|x-y|}\left\vert \frac{x-y}{|x-y|}-%
\frac{x}{|x|}\right\vert |v(y)|\left\vert \frac{y}{|y|}-\frac{x-y}{|x-y|}%
\right\vert \frac{e^{-|y|}}{|y|}dy &=& \\
\int_{|y|<2|x|/3}\frac{e^{-|x-y|}}{|x-y|}\left\vert \frac{y}{|y|}-\frac{x}{%
|x|}\right\vert |v(x-y)|\left\vert \frac{y}{|y|}-\frac{x-y}{|x-y|}%
\right\vert \frac{e^{-|y|}}{|y|}dy && \\
<\frac{Ce^{-|x|}}{|x|^{1.5+\varepsilon }+1}\int\limits_{0}^{|x|}\frac{d\rho
}{\rho +1}<\frac{Ce^{-|x|}}{|x|^{1.5}+1} &&
\end{eqnarray*}
$\blacksquare$

Proof of the theorem \ref{th7}. Let us iterate the second
resolvent identity to get the needed estimate for $G_{i}(x,0)$
\begin{equation*}
G_{i}(x,0)=G_{i}^{0 }(x,0)-\int\limits_{\mathbb{R}^{3}}G_{i}^{0
}(x,s)\beta v(s)G_{i}(s,0)ds
\end{equation*}%
Using lemmas \ref{dwarf}--\ref{goblin} and explicit formula for
$G_{i}^{0 }(x,s)$, we see
that the $n-$th term in the corresponding series has the following form%
\begin{equation}
\frac{e^{-|x|}}{|x|+1}\left[ \left( \frac{\alpha \cdot
x}{|x|}+1\right) \varphi _{1}^{(n)}(x)+(|x|+1)^{-0.5}\varphi
_{2}^{(n)}(x)\right] \label{knew}
\end{equation}%
with $|\varphi_{1(2)}^{(n)}(x)|<[C(\varepsilon )C_{v}]^{n}$. That
can be easily proved by the induction. Indeed, for the first term,
we have the representation (\ref{fri}). Assume that (\ref{knew})
holds for the $n$-th term $T_n$. Then,
\[
T_{n+1}(x)=-\int\limits_{\mathbb{R}^{3}}G_{i}^{0 }(x,s)\beta
v(s)T_n(s)ds=-(I_1+I_2+I_3+I_4)
\]
where
\[
I_1(x)= \frac{i}{4\pi} \int\limits_{\mathbb{R}^3}
\frac{\exp(-|x-s|)}{|x-s|} \left(   \frac{\alpha \cdot
(x-s)}{|x-s|}+1 \right) \beta v(s) \left(  \frac{\alpha \cdot
s}{|s|}+1 \right)\frac{\exp(-|s|)}{|s|+1}\, \varphi_1^{(n)}(s)ds
\]
\[
I_2(x)= \frac{i}{4\pi} \int\limits_{\mathbb{R}^3}
\frac{\exp(-|x-s|)}{|x-s|} \left(   \frac{\alpha \cdot
(x-s)}{|x-s|}+1 \right) \beta v(s) \left(
\frac{\exp(-|s|)}{(|s|+1)^{1.5}} \right) \varphi_2^{(n)}(s)ds
\]
\[
I_3(x)=\frac{i}{4\pi} \int\limits_{\mathbb{R}^3} \frac{\alpha
\cdot (x-s)}{|x-s|^2} \left( \frac{\exp(-|x-s|)}{|x-s|}\right)
\beta v(s) \left( \frac{\alpha \cdot s}{|s|}+1
\right)\frac{\exp(-|s|)}{|s|+1}\, \varphi_1^{(n)}(s)ds
\]
\[
I_4(x)=\frac{i}{4\pi} \int\limits_{\mathbb{R}^3} \frac{\alpha
\cdot (x-s)}{|x-s|^2} \left( \frac{\exp(-|x-s|)}{|x-s|}\right)
\beta v(s) \left( \frac{\exp(-|s|)}{(|s|+1)^{1.5}} \right)
\varphi_2^{(n)}(s) ds
\]
We apply lemma \ref{goblin}  to $I_1$, lemma \ref{ork} to $I_2$,
and lemma \ref{dwarf} to $I_3$ and $I_4$. If $C_{v}$ is small
enough, we will get uniform convergence of the series and the
estimate (\ref{numer}). Clearly, $\|\cal{P}_1-1\|<\delta$ with
$\delta\to 0$ if $C_v \to 0$.
 $\blacksquare
$

{\bf Remark.} It is clear that the asymptotics of the Green's
function $G_i(x,s), |s|<1$ can be obtained similarly: it is close
to $G_i^0(x,s)$ as $|x|\to\infty$.

Fix any $\varepsilon >0$. Consider the Dirac operator $\cal{H}$ with $%
|v(x)|<C_{v}(|x|+1)^{-0.5-\varepsilon }$, where $C_{v}$ is sufficiently
small, and operator $\cal{H}_{s}$ on $\mathbb{R}^{3}$ with $v(x)=0$ on $%
|x|<1 $. Then, for any nonzero function $f(x)=(f_{1},f_{2},f_{3},f_{4})^{t}%
\in \left[ L^{2}(\mathbb{R}^{3})\right] ^{4}$ with support in the
unit ball, the function
\[
h(x)=\int\limits_{\mathbb{R}^{3}}G_{i}(x,s)f(s)ds
\]
satisfies equation $Hh=ih$ ($h$ does not satisfy boundary
condition on $\Sigma$ unless $h=0$ in $\Omega$). Therefore, by the
theorem \ref{th7} and by the remark above, we can control the
``amplitude" of $h$. For $|x|$ large enough,

\begin{equation*}
e^{|x|}|x|h(x)=\frac{i}{4\pi}\left( \frac{\alpha \cdot x}{|x|}%
+1\right) \int\limits_{\mathbb{R}^{3}}\exp \langle
\frac{x}{|x|},s\rangle f(s)ds+\hat{f}(x)
\end{equation*}%
and%
\begin{equation*}
\|\hat{f}(x)\|<\delta ||f||_{1}
\end{equation*}%
with $\delta\to 0$ as $C_v\to 0$. For the integrals, we have
\begin{equation*}
\int\limits_{\mathbb{R}^{3}}\exp \langle \frac{x}{|x|},s\rangle |f_{j}(s)|ds>e^{-1}\int%
\limits_{\mathbb{R}^{3}}|f_{j}(s)| ds,(j=1,\ldots ,4)
\end{equation*}
The ``amplitude" is calculated as
\[
\cal{A}(|x|^{-1}x) =\lim_{|x|\to \infty} |x|\exp(|x|-1)
(h_3(x),h_4(x))^t
\]
Take $f_1=f_2=0$ and $f_3, f_4$ -- arbitrary non-negative. Then,
$\cal{A}$ is well-defined as an element of $[L^2(\Sigma)]^2$ for
$v_n=v(x)\chi_{|x|<n}(x)$. Denote this sequence by $\cal{A}_n$.
Now, theorem \ref{th7} says that $\cal{A}_n$ does not go to zero.
Using the reduction to a one-dimensional system and
(\ref{paradigma}), one can show that the spectral measure of an
element $f$ has an a.c. component which fills $\mathbb{R}$.

It is likely, that more careful estimates might prove some analog
of theorem \ref{th7} for $v$ satisfying condition
(\ref{condition}). One of the main ideas of the paper is as
follows. Take the spectral parameter as far from the spectrum as
we wish. If we manage to find at least one solution to homogeneous
problem that satisfies certain asymptotics at infinity with
nonzero ``amplitude", then an a.c. spectrum fills the whole line.
The main advantage is that we need to do some rather delicate
analysis of, say, Green's function far from the spectrum, not on
the real line. That is an interesting problem to study the Green's
function for operator (\ref{dirac8}) when the potential is random
with slow decay or with no decay at all (the Anderson model). The
approach of theorem \ref{th1} does not work in this case. In the
meantime, analysis done in \cite{Bour} together with estimates
used in the proof of theorem \ref{th7} seem to be relevant for the
problem.

\begin{center}
\textbf{Acknowledgment }
\end{center}

The author is grateful to  Ari Laptev, Oleg Safronov, and Barry
Simon for the useful discussions of the results obtained. This
work was partially supported by the NSF grant INT-0204308. Some of
the results were obtained during the author's visit to KTH,
Stockholm in Spring, 2003.

\end{document}